# 29 W High Power CW Supercontinuum Source

**B.A. Cumberland, J.C. Travers, S.V. Popov and J.R. Taylor**

Femtosecond Optics Group, Imperial College London, London, SW7 2AZ, United Kingdom

http://www.femto.ph.imperial.ac.uk

**Abstract:** A 29 W CW supercontinuum spanning from 1.06 to 1.67 µm is generated in a short length of PCF with two zero dispersion wavelengths. The continuum has the highest spectral power density, greater than 50 mW/nm up to 1.4 µm, reported to date. The use of a short length of PCF enables the continuum to expand beyond the water loss at 1.4 µm. The dynamics of the continuum evolution are studied experimentally and numerically with close attention given to the effects of the water loss and the second zero dispersion wavelength.

## 1. Introduction

Since the advent of photonic crystal fibers (PCF) [1] there has been renewed activity in the development and study of supercontinuum sources due to the increasing number of applications for such sources [2-6] and the flexible control offered by PCF. A wide variety of pump sources including femtosecond, picosecond, nanosecond and CW sources at a variety of wavelengths [7], or in some cases multiple wavelengths [8] have been demonstrated. Combined with a fairly extensive exploration of fiber structures including: varying zero dispersion wavelengths [7], fibers with two zero dispersion wavelengths [9], long tapers generating continua spanning the entire transmission window of silica [10], alternative glass compositions enhancing short wavelength generation [11] and non-silica based fibers [12, 13] have mapped out the field fairly extensively.

While much of the underlying physics has been known since the 1980s more recent theoretical work has revisited and reexamined the variety of nonlinear interactions in these structures under a variety of pump conditions [7, 14-16]. It may appear that work in this field is close to complete but there are many challenges remaining to be addressed. Some examples include, the generation of mid-IR sources, improvements in the spectral bandwidth in the CW regime, a shift to sources that are flat on a linear scale and improved spectral power densities to name a few.

In this work we enhance the spectral power densities available and study the dynamics of CW continuum formation in a double zero PCF while extending the bandwidth. Supercontinua based upon CW pump sources have offered the highest spectral power densities (SPD) to date and have been demonstrated using 1 µm and 1.55 µm pumps [9, 17-22], with the highest being based around a 1.56 µm ASE pump and 300 m of highly nonlinear fiber providing a continuum with a SPD of 16 mW/nm from 1.59 to 1.98 µm with a 0.7 dB flatness [20]. Continua using 1 µm based pumps have traditionally been spectrally limited by the high water losses in PCF at 1.38 µm [9], though Travers *et al.* demonstrated extension beyond 1.38 µm using low water loss PCFs [21]. In this work we demonstrate negation of the high water loss by using a short length of PCF, which also aids the increase in SPD. Ultimately we generate a supercontinuum spanning from 1.06 to 1.67 µm with 29 W of output power utilizing a 50 W Yb fiber pump laser. This corresponds to a spectral bandwidth of 600 nm at 8 dB and a power density of more than 50 mW/nm up to 1.4 µm. A similar idea has recently been explored theoretically by Mussot *et al.* [23] and where appropriate we will draw comparisons from their work.

The rest of this paper has been split into several sections. In the next section we will discuss the fiber, its properties and how to achieve low loss splices in order to enable high power pumping. In section 3 we will outline the experiment before going on to discuss our simulations and results in section 4. Finally we will conclude in section 5.

## 2. The Fiber

A 20 m double zero PCF was used as the basis of this work. A scanning electron microscope (SEM) image was taken of one end of the fiber and used to calculate the dispersion, nonlinearity and mode field diameter by modeling a step index fiber using a vectorial effective index method for the cladding [24]. In order to perform this calculation the pitch ($\Lambda$) and hole diameter (d) of the PCF are required and the cladding structure is assumed to be uniform. By examination of the SEM images it is clear that the hole diameter varies significantly and non-uniformly across the structure. Applying these values to the calculation leads to uncertainty in the dispersion profile especially with regards to the second zero dispersion wavelength (ZDW). In order to reduce this, the group velocity delay was measured between 1.5 and 1.6 µm. From this the dispersion was calculated and the complete dispersion curve was fitted to this small subsection of real data by keeping the pitch fixed and varying the hole diameter. Using this technique the ZDWs were calculated as 0.81 µm and 1.73 µm. The complete dispersion curve is shown in figure 1a along with the nonlinearity in figure 1b and an inset of an SEM of PCF in figure 1b. At the pump wavelength of 1.07 µm the mode field diameter (MFD) is 2.2 µm with a dispersion of 65 ps/(nm km) and a nonlinearity of 0.043 (W m)$^{-1}$.

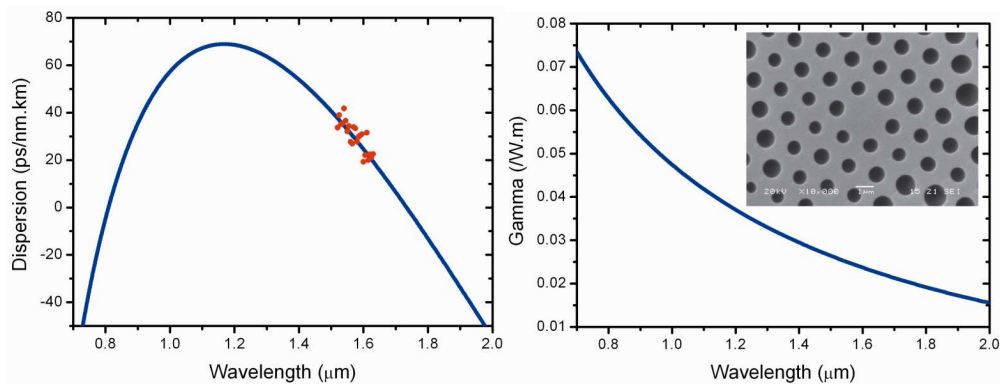

Figure 1. (a) The calculated dispersion curve (solid line) and experimental measurements (circles) (b) The calculated nonlinearity. Inset an SEM of the fiber.

Typically PCFs are difficult to splice to owing to their micro-structured nature and small MFDs. Over the last few years several papers have been published on splicing PCFs to standard fibers [25-27] with good results. Traditionally high splice losses of several dB occur due to mode field mismatch and uncontrolled collapse of the micro-structure leading to high waveguiding losses. In our experience there is no single set of fusion parameters that can be applied universally. As a result the splice conditions need to be individually optimized for every PCF. In general we find that larger core PCFs (4 µm and above) can be spliced directly to a standard fiber similar to Flexcore with losses as low as 0.3 dB and typically of the order of 0.5-0.75 dB. These splices are often achieved using a long low power fusion arc with the electrodes offset to the side of the standard fiber. The offset is generally quite small and helps prevent hole collapse in the PCF by reducing the exposure of the PCF to the arc while maintaining a strong enough arc to melt the solid core fiber. For small core PCFs, as used in this work, an intermediated single mode high NA fiber (such as Nufern Ultra High NA fiber) allows for improved mode field matching between Flexcore and the PCF. In this regime the high NA fiber is splice to the Flexcore using a mode field matching technique and the PCF is spliced with a very short, offset, arc to the high NA fiber.

An Ericsson fusion splicer (FSU 975-PM-A) was used and typical losses of 0.15-0.30 dB were achieved for the high NA to Flexcore splice and losses as low as 0.4 dB for the PCF to high NA fiber. Thus the total loss between the Flexcore and the PCF was typically 0.55-1.00 dB. It should be noted that the quality of the cleave, cleave angle, fusion current, duration and electrode offset are critical in achieving low loss splices.

Even with such low splice losses further precautions need to be taken. Optical glue is used to secure the splice and more importantly strip and scatter stray light at the point were the fiber coating begins. Failure to do this can often result in burning of the coating, as stray cladding modes created at the splice couple in to it. Additionally the entire splice needs to be thermally managed as heat build up may begin to cause splice degradation. We believe that localized heating can result in slight distortion of the splice interface resulting in increased loss thus further heating and eventually resulting in significant loss and often a fiber fuse [28, 29]. Finally the output of the PCF was angle-cleaved to reduce the reflection from the end facet. Such reflections, from the output of the fiber and poor quality input splices, significantly enhance any Raman Stokes lines in the continuum and hence should be minimized in order to improve continuum flatness.

## 3. Experimental Setup

A 50W CW Yb fiber laser (IPG Photonics) was spliced to 20 m of the above PCF via an intermediate high NA fiber (Nufern) as described above with a total splice loss of 0.6 dB. The output end of the PCF was angle cleaved and either coupled into an optical spectrum analyzer (Advantest Q8384, Anritsu MS9710B) via an intermediate fiber or collimated and launched into an automated Spex 500 spectrometer in combination with a PbS IR detector and lock-in amplifier for spectral measurements. Output power was measured on a thermal power meter head (Molectron). The average pulse duration was measured by spectrally slicing the supercontinuum using a Spex-minimate 0.25 m monochromator followed by re-collimation and launch into an Inrad autocorrelator (Model 5-14b). In order to reduce the average power on the monochromator's gratings several high reflector mirrors were used in reflection to perform an initial broad spectral slice. The autocorrelations were fitted assuming a sech[2] pulse shape. An outline of the experimental setup is shown in figure 2.

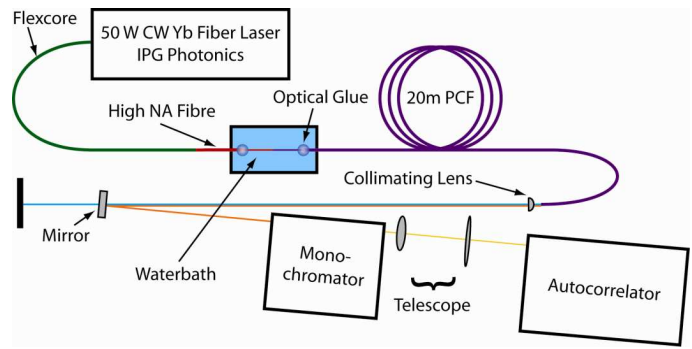

Figure 2. Experimental setup.

## 4. Results and Simulations

The PCF was pumped at a variety of pump powers, under varying conditions and a series of measurements were made before a cutback of the fiber was performed. In this section we will present and discuss the results as well as using theory and simulation to analyze the results.

### 4.1 The Numerical Model

The results were modeled by solving the general nonlinear Schrödinger equation using a fourth order Runge-Kutta in the interaction picture method [30]. Our model accounts for the dispersion up to arbitrary orders, the dispersion of the nonlinearity, the Raman effect and the spectrally dependent loss profile of the fiber, including the water loss (experimentally measured as 650 dB/km at 1.38 μm). The initial conditions, corresponding to our laser, were obtained by adding a random spectral phase to each frequency bin of the average spectrum similar to that used by Vanholsbeeck *et al.* [31] and Barviau *et al.* [32]. To simulate CW continuum formation within a reasonable computation time we modelled a time window of only 256 ps, however, we note that all of the physical processes involved in the spectral broadening act over much shorter time scales. The grid contained $2^{17}$ points, sufficient to cover our frequency window, and we used an adaptive spatial step size to maintain accuracy [33]. In order to produce a spectral output similar to that seen in real life the model was run 12 times with a different random spectral phase each time. The results in this paper are therefore the averaged results unless otherwise stated.

### 4.2 Experimental Results

At full pump power (50 W) 44 W is launched into the PCF producing a 29 W supercontinuum with an 8 dB bandwidth of 600 nm (1.06 to 1.67 μm) as shown in figure 3a. This corresponds to a SPD above 50 mW/nm between the pump and 1.38 μm as shown in figure 3b. The continuum is formed by modulation instability producing fundamental solitons, which then undergo soliton-self-frequency-shift to longer wavelengths, before generating dispersive waves beyond the second zero dispersion of the PCF. The output power and spectral flatness are slightly curtailed by the high water loss at 1.38 μm in the fiber (13 dB in the 20 m length) whilst the supercontinuum's bandwidth is ultimately limited by the second zero dispersion wavelength at 1.73 μm.

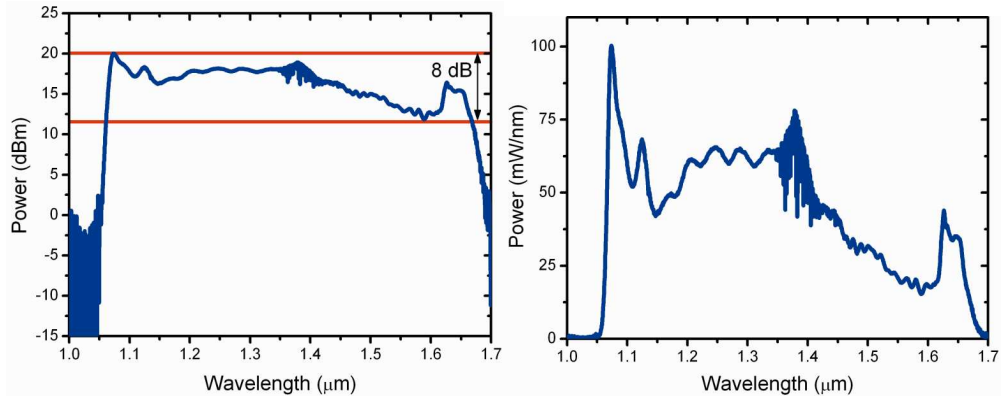

Figure 3. Supercontinuum generated by 44 W launched into the PCF on a log scale (a) and a linear scale (b)

At lower pump powers the classic spectral wings from modulation instability are clearly visible as can be seen in figure 4a. It can be shown that the frequency for maximum MI gain is given by $\Omega_{max} = \pm(2\gamma P_0 / |\beta_2|)^{1/2}$ where $\Omega_{max}$ is the frequency shift for maximum gain, $\gamma$ is the nonlinearity, $P_0$ is the peak power and $\beta_2$ is the group velocity dispersion (GVD) parameter [34]. For this fiber at the pump wavelength $\gamma = 0.043$ (W m)$^{-1}$, $\beta_2 = -0.040$ ps$^2$/m and for a pump power of $P_0 = 1.34$ W we find $\Omega_{max} = 1.7$ THz. This corresponds to a wavelength change of 1 nm from the pump, which is slightly less than that seen in figure 4a (solid green line). This would perhaps indicate that the real value of $\beta_2$ is lower.

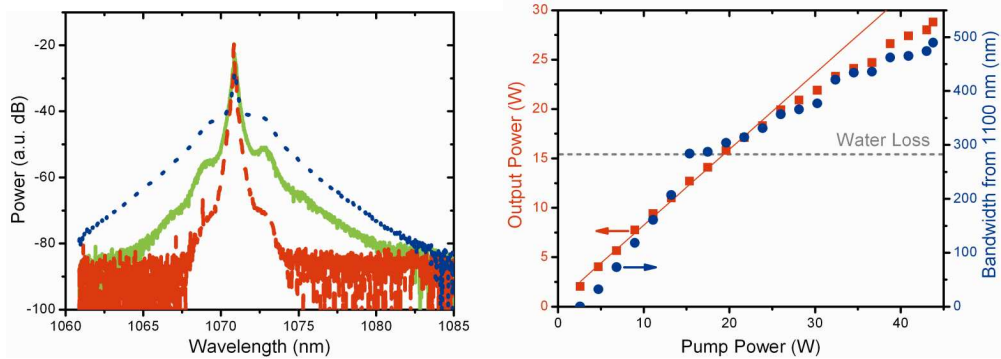

Figure 4. (a) Modulation instability sidebands at 1.1 W (red dashed), 1.3 W (green solid) and 1.9 W (blue dot). (b) Evolution of output power (squares) and bandwidth (circles) with pump power

The evolution of the continuum with pump power is shown in figure 4b. Here we can see progress of the continuums output power along with the increase in bandwidth of the continuum, where the bandwidth is measured from 1100 nm to the point where the power has dropped by a further 5 dB. It is clear that the rate of increase of the continuum bandwidth is curtailed once the water loss at 1.38 μm is reached (marked on the graph). It also shows that the power growth falls off from a linear fit as the continuum broadens beyond 1.38 μm. While the use of such a short length of PCF has enabled us to extend beyond the water loss, in contrast with earlier results [9], it is clear that the water loss still affects the continuum contrary to a more optimistic analysis by Mussot *et al.* [23]. Several things should be noted however, Mussot *et al.* simulated several double zero fibers with the longest second ZDW at 1.31 μm, still short of the water loss. Additionally we experimentally measured the fiber loss spectrum and found a low loss of 10 dB/km on the short wavelength side of the water loss but a much higher 60 dB/km on the long wavelength side. It is likely that this higher loss on the long wavelength side contributes to a reduction in bandwidth growth and a steady fall in the SPD beyond 1.38 μm as seen in figure 3b.

The efficiency of the soliton self-frequency shift (SSFS) is clearly an important factor for the generation of the continuum. The soliton energy ($E_s$) is given by $E_s = 2|\beta_2|/\gamma\tau_0$ where $\tau_0$ is the soliton duration. For a double zero fiber, when pumping relatively far from the first ZDW the relationship of $\beta_2/\gamma$ is fairly fixed as the wavelength is increased. This means that $\tau_0$ remains consistently short enabling efficient SSFS. Contrasting this with a PCF having only a single ZDW we find that $\beta_2/\gamma$ generally increases resulting in longer duration solitons and thus a loss of efficiency in the SSFS. As the solitons evolve from the MI we can estimate the soliton duration by calculating the frequency of maximum MI gain when $P_0 = 44$ W. From this we find $\Omega_{max} = 9.73$ THz which gives a period of $t_m = 0.65$ ps for the pulse train. Assuming that the full width half max of any soliton formed does not exceed $t_m/2$ and accounting for the sech$^2$ pulse shape of a soliton we find that the maximum duration of a fundamental soliton would be $\tau_0 = 211$ fs. This is slightly shorter than measured (figure 5a). From figure 5a it can be seen that the pulse duration decreases with a minimum duration at 1.14 μm and then steadily increases with wavelength. This may be explained by Raman amplification of the solitons followed by a slow loss of energy at longer wavelengths. Also included in figure 5a is a numerical autocorrelation from a single shot of the simulation, showing reasonable agreement with the experimental results.

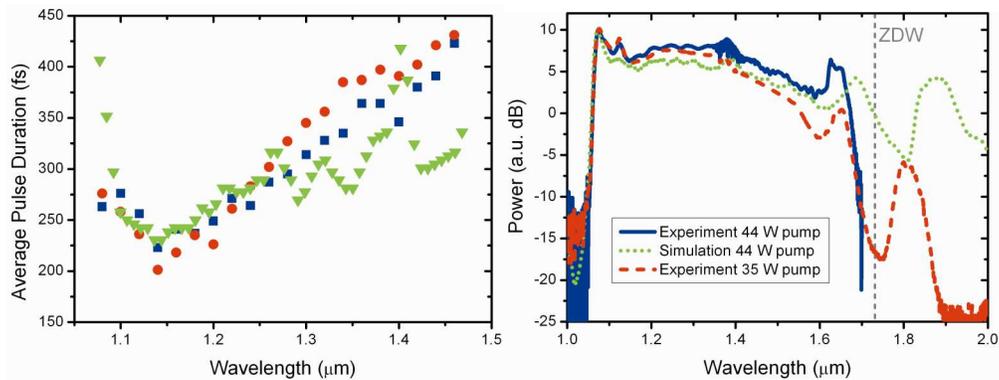

Figure 5. (a) Average soliton duration with wavelength for two sets of measurements (circles and squares) along with a numerical autocorrelation of a single shot of the simulation (triangles) (b) Experimental and numerical continuum results, normalized.

The downside of using a double zero PCF is that the solitons can not travel beyond the second ZDW which limits the long wavelength edge of the continuum. From theory [15] it is expected that the soliton self-frequency shift will undergo cancellation via spectral recoil and the generation of a dispersive wave beyond the second ZDW. This can be seen in figure 5b which shows the spectral output from an experimental measurement of the spectrum further into the infrared with 35 W launched in the PCF. For comparison the result of a numerical simulation and experiment with 44 W launched into the PCF is shown. In all three traces an increase in the SPD can be seen just before the ZDW. This is due to the cancellation of the SSFS leading to a build up of solitons and hence energy just before the ZDW. On the long wavelength side of the ZDW a dispersive wave can be seen. Notably there is a difference in the location of the soliton cancellation and the dispersive wave in the numerical result compared to the experimental measurements. There are several explanations for this including: inaccuracy in our calculated dispersion curve, requiring a slightly shorter ZDW; the possibility of the solitons in our model having higher energies (shorter durations as figure 5a) and therefore phase matching to a dispersive wave at a longer wavelength. The other interesting thing to note is that the dispersive wave sees a much greater amplification in the numerical simulation than our experimental measurement. This may simply be due to the lower pump power of the experiment versus the simulation or perhaps the fiber loss is greater beyond 1.7 μm than what we used in the model. The only remaining difference between the simulation and experiment is the presence of a Raman Stokes peak in the experimental data, caused by a residual weak back reflections between the angle cleaved end of the PCF and the splice. Finely tuning the cleave angle allows the Stokes line to be minimized. Splicing a multimode fiber to the end of the PCF is expected to eliminate it.

The soliton cancellation and dispersive wave generation is clearly seen if we examine a spectrogram (i.e. XFROG) of the supercontinuum generated by our numerical model as shown in figure 6. The spectrogram is from a single shot rather than an average of 12 simulations as discussed in section 4.1. The spectrogram shows the development of solitons from MI which then shift to longer wavelengths, colliding and shedding energy before coming to rest just before the second ZDW. Beyond the second ZDW the corresponding dispersive waves can be seen forming. It is also clear from the spectrum (on the right side of the spectrogram) that the water loss at 1.38 μm reduces the SPD of the continuum and that the soliton cancellation results in an increased SPD just before the second ZDW.

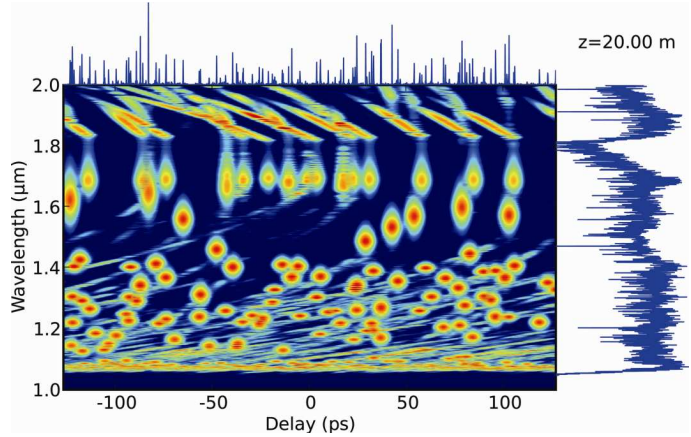

Figure 6. Spectrogram from a single shot numerical simulation for 20 m of PCF and $P_0$= 44 W

Finally the evolution of the continuum with length was measured via a cutback of the PCF. This is shown in figure 7a. We can see that the continuum does not significantly broaden beyond 1.4 μm until it has propagated though 14 m. Note the modulations visible at the long wavelengths beyond 15 m are due to a burnt patch cable, which was replaced during the cutback. It is also worth noting that several poor angle cleaves result in the temporary appearance of tHigh Power Continuumhe first Raman Stokes line. Comparing with the numerical simulation in figure 7b we see the numerical simulation broadens out much faster in the first 5 m and broadens beyond 1.4 μm 3 m earlier than the experiment. It is also clear that it is a smaller number of high energy solitons that are crossing the water loss before bunching up due to SSFS cancellation. The more rapid broadening of the bandwidth can be explained by assuming that the random spectral phase components which are added to simulate the laser result in intensity fluctuations in the time domain which are greater than that seen in reality, in fact these fluctuations are dependent on the numerical grid chosen. In order to achieve excellent agreement between simulation and reality a more accurate model of the pump laser would be required.

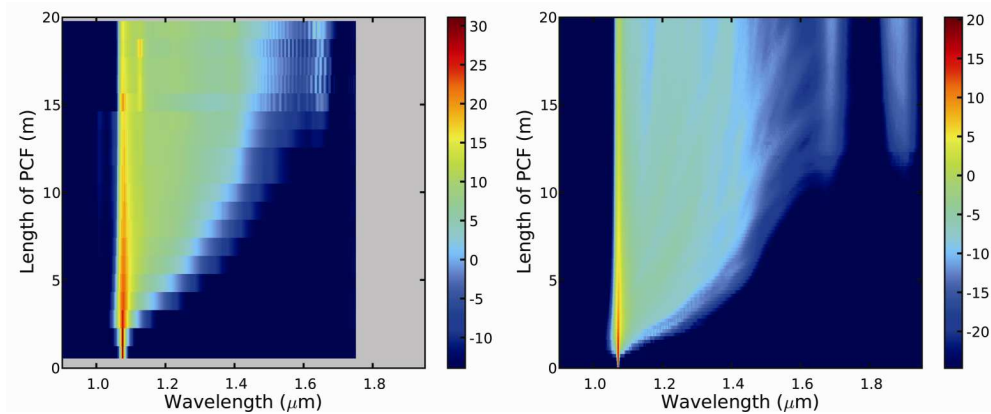

Figure 7. (a) Experimental measurement of continuum evolution via a cutback of the PCF (b) Numerical simulation of continuum evolution with length of PCF

## 5. Conclusion

In this work we have generated a high power CW supercontinuum with a bandwidth extending from 1.06 to 1.67 μm at the 8 dB level. The continuum has the highest spectral power density demonstrated to date, producing more than 50 mW/nm up to 1.4 μm. This has been made possible by significantly reducing the splice loss between a standard fiber and a PCF in order to launch high powers into the PCF. Although the use of a short length of PCF reduced the water loss at 1.38 μm significantly and allowed the continuum to extend beyond it, the loss still plays a dominant role in the continuum evolution. A short low water loss PCF should enable the generation of a flatter supercontinuum beyond 1.38 μm. Similarly the use of a single zero dispersion wavelength fiber will enable continued extension of available bandwidth on the infra-red side of the pump. It is expected however that the continuum formation dynamics will be affected by the increasing ratio of GVD over the nonlinearity. In a single ZDW PCF this may act to contain the long wavelength edge of the continuum. The use of a double zero dispersion wavelength fiber in this work has experimentally and numerically confirmed some interesting physical processes including dispersive wave generation and soliton recoil in a CW supercontinuum. It may also prove to be a useful tool to control the long wavelength edge of the continuum when needed for specific applications.